\documentclass[twocolumn,showpacs,preprintnumbers,amsmath,amssymb,superscriptaddress]{revtex4}

\usepackage{graphicx}
\usepackage{graphicx,epsf} 
\usepackage{dcolumn}
\usepackage{bm}
\usepackage{latexsym}
\newcommand{\kav}{\langle k \rangle}
\newcommand{\xsize}{\epsfxsize=7.0cm}

\begin{document}

\title{Optimization of Network Robustness to Random Breakdowns}

\author{Gerald~Paul}
\affiliation{Center for Polymer Studies and Dept.\ of Physics, Boston
  University, Boston, MA 02215, USA} 
\email{gerryp@bu.edu}

\author{Sameet Sreenivasan}
\affiliation{Center for Polymer Studies and Dept.\ of Physics, Boston
  University, Boston, MA 02215, USA} 

\author{Shlomo Havlin}
\affiliation{Center for Polymer Studies and Dept.\ of Physics, Boston
  University, Boston, MA 02215, USA} 
\affiliation{ Minerva Center and Department of Physics, Bar Ilan
University, Ramat Gan 52900, Israel}

\author{H.~Eugene Stanley}
\affiliation{Center for Polymer Studies and Dept.\ of Physics, Boston
  University, Boston, MA 02215, USA} 
 
\pacs{89.20.Hh,02.50.Cw,64.60.Ak}

\begin{abstract}

We study network configurations that provide optimal robustness to
random breakdowns for networks with a given number of nodes $N$ and a
given cost---which we take as the average number of connections per node
$\kav$. We find that the network design that maximizes $f_c$, the
fraction of nodes that are randomly removed before global connectivity
is lost, consists of $q=[(\kav-1)/\sqrt\kav]\sqrt N$ high degree nodes
(``hubs'') of degree $\sqrt{\kav N}$ and $N-q$ nodes of degree 1. Also,
we show that $1-f_c$ approaches 0 as $1/\sqrt N$---faster than any other
network configuration including scale-free networks.  We offer a simple
heuristic argument to explain our results.

\end{abstract}  

\maketitle

\section{Introduction}

Recently there has been much interest in determining network
configurations which are robust against various types of attacks
\cite{Albert,Paxon,Cohen2000,Callaway,Cohen2001,Cohen2002,Valente2004,Paul2004,Tanizawa}.
While there have been studies of complex combinations of different types
of attacks, surprisingly, there has been no focused analysis of the the
elementary case of robustness against simple random breakdowns or
attacks.

We first study simple random networks.  Simple networks contain no self
loops or multiple edges neither of which add to the robustness of a
network to random removal of nodes.  For simple random networks we can
determine the optimal network configuration analytically.  Randomly
constructed networks, however, may have disconnected components, so we
also consider networks constructed in a way that ensures they consist
initially of a single cluster of connected nodes.  These networks are
degree correlated. For degree correlated networks, there currently exist
no closed-form expressions with which we can determine $f_c$
analytically, so we study them using Monte-Carlo simulations.  We find
that the optimal configuration for both the randomly constructed and the
degree correlated networks consists of $q \sim \sqrt{N}$ high degree
nodes ({\it hub\/} nodes) of degree $k_2 \sim \sqrt{ N}$ and $N-q$ nodes
of degree 1 ({\it leaf\/} nodes).

\section{Uncorrelated Networks}

\subsection{Theory}

We first treat simple random networks.  It is known
\cite{Chung,Burda,Boguna,Catanzaro} that for any desired random degree
distribution, simple networks can be created only if $P(k)=0$ for $k$
greater than the {\it structural cutoff}
\begin{equation}
K_s \equiv \sqrt{ \kav N}
\label{KcStruct}
\end{equation}
So we must limit our networks to those with maximum degree less than
$\sqrt{\kav N}$.   For networks with this constraint we can use the equation
\cite{Cohen2000,Paul2005}
\begin{equation}
f_c=1-{1 \over{\kappa-1}}
\label{fc}
\end{equation}
where
\begin{equation}
\kappa \equiv \frac{ \langle k^2 \rangle}{\kav}
\label{kappa}
\end{equation}
to determine $f_c$. 

Since we fix $\kav$, maximizing $f_c$ is equivalent to maximizing
$\langle k^2 \rangle$.  We must maximize
\begin{equation}
h(P) \equiv \sum_{k=1}^{K_s} k^2  P(k)
\label{ksq}
\end{equation} 
under the following constraints
\begin{equation}
P(k) \ge 0
\label{Pk0}
\end{equation}
\begin{equation}
\sum_{k=1}^{K_s} k P(k)=\kav.
\label{kavconstraint}
\end{equation}
\begin{equation}
\sum_{k=1}^{K_s} P(k)=1
\label{norm}
\end{equation}
We first show that there can be no more than two unique values of $k$ at
which $P(k)$ is non-zero if $h(P)$ is to be maximized. Assume that there
are $m>2$ non-zero values $P(k_1), P(k_2), P(k_3) . . . P(k_m)$ needed
to maximize $h(P)$. Using the method of Lagrange
multipliers~\cite{Arfken} we can write
\begin{eqnarray}
\frac{\partial\left(\sum_{i=1}^{K_s} k_i^2 P(k_i)\right)}{\partial P(k_j)} + \lambda_1
\frac{\partial\left(\sum_{i=1}^{K_s} k_i P(k_i)-\kav\right)}{\partial
  P(k_j)}  \notag \\
+\lambda_2 
\frac{\partial\left(\sum_{i=1}^{K_s}  P(k_i)\right)}{\partial P(k_j)}=0 
\label{lagrange1}
\end{eqnarray}
or 
\begin{equation}
k_j^2  + \lambda_1 k_j +  \lambda_2 =0 \qquad [1 \le j \le m]
\label{lagrange2}
\end{equation}
where $\lambda_1$ and $\lambda_2$ are constants.  Solving
(\ref{lagrange2}) we find at most only two unique solutions for the
values of $k_j$.

Analyzing the problem now with only two values $k_1$ and $k_2$ for which
$P(k)$ are non-zero, we find that $h(P)$ is maximized when $k_1$ and
$k_2$ take on the boundary values
\begin{subequations}
\label{k1k2}
\begin{eqnarray}
k_1=1      \\
k_2=K_s.
\end{eqnarray}
\end{subequations}  
and
\begin{subequations}
\label{Pk1Pk2}
\begin{eqnarray}
P(k_1)=1-\frac{\kav-1}{\sqrt{\kav }} \frac{1}{\sqrt{N}} \\
P(k_2)=\frac{\kav-1}{\sqrt{\kav }} \frac{1}{\sqrt{N}}
\label{Pk2}
\end{eqnarray}
\end{subequations}  
For these values $1-f_c$ assumes its minimal value
\begin{equation}
(1-f_c)_{\rm min}=\frac{\kav \sqrt{\kav N}} { (\kav-1) (1-\kav N +\sqrt{\kav N})}
\label{fcmin}
\end{equation}
For large $N$,
\begin{equation}
(1-f_c)_{\rm min} \sim \frac{\sqrt{\kav}}{(\kav-1)} \frac{1}{\sqrt{N}}.
\label{fcminN}
\end{equation}

\subsection{Simulations}

We next perform Monte Carlo simulations to test the results found above.
We consider the degree distribution that represents a network of $q$ hub
nodes and $N-q$ leaf nodes,
\label{pkxx}
\begin{equation}
P(k)=
\begin{cases}
{N-q \over N} & k=1 \\
  {q \over N} & k=k_2 \\
           0  & \mbox{otherwise}, 
\end{cases} 
\label{pkx}       
\end{equation}
where
\begin{equation}
k_2={(\kav-1) N + q \over{q}}.
\label{k2}
\end{equation}
Our aim is to find the value of $q$ which maximizes the robustness of
the network.

We create networks using the method described in Ref.~\cite{Molloy}.  We
then randomly delete nodes in the network and after each node is
removed, we calculate $\kappa$. We use the criterion
\begin{equation}
\kappa < 2
\label{kappa2}
\end{equation}
for loss of global connectivity
\cite{Molloy,Cohen2000,Callaway,Cohen2002}.  When $\kappa$ becomes less
than 2 we record the number of nodes $n_r$ removed up to that point.  This
process is performed for many realizations of random graphs with 
the degree distribution of Eq.~(\ref{pkx}) and, for each graph, for many different
realizations of the sequence of random node removals. The threshold
$f_c$ is defined as
\begin{equation}
f_c \equiv \frac{ \langle n_r \rangle}{N}
\label{fcdef}
\end{equation}
where $\langle n_r \rangle$ is the average value of $n_r$.

\begin{figure}
\centerline{
\xsize
\epsfclipon
\epsfbox{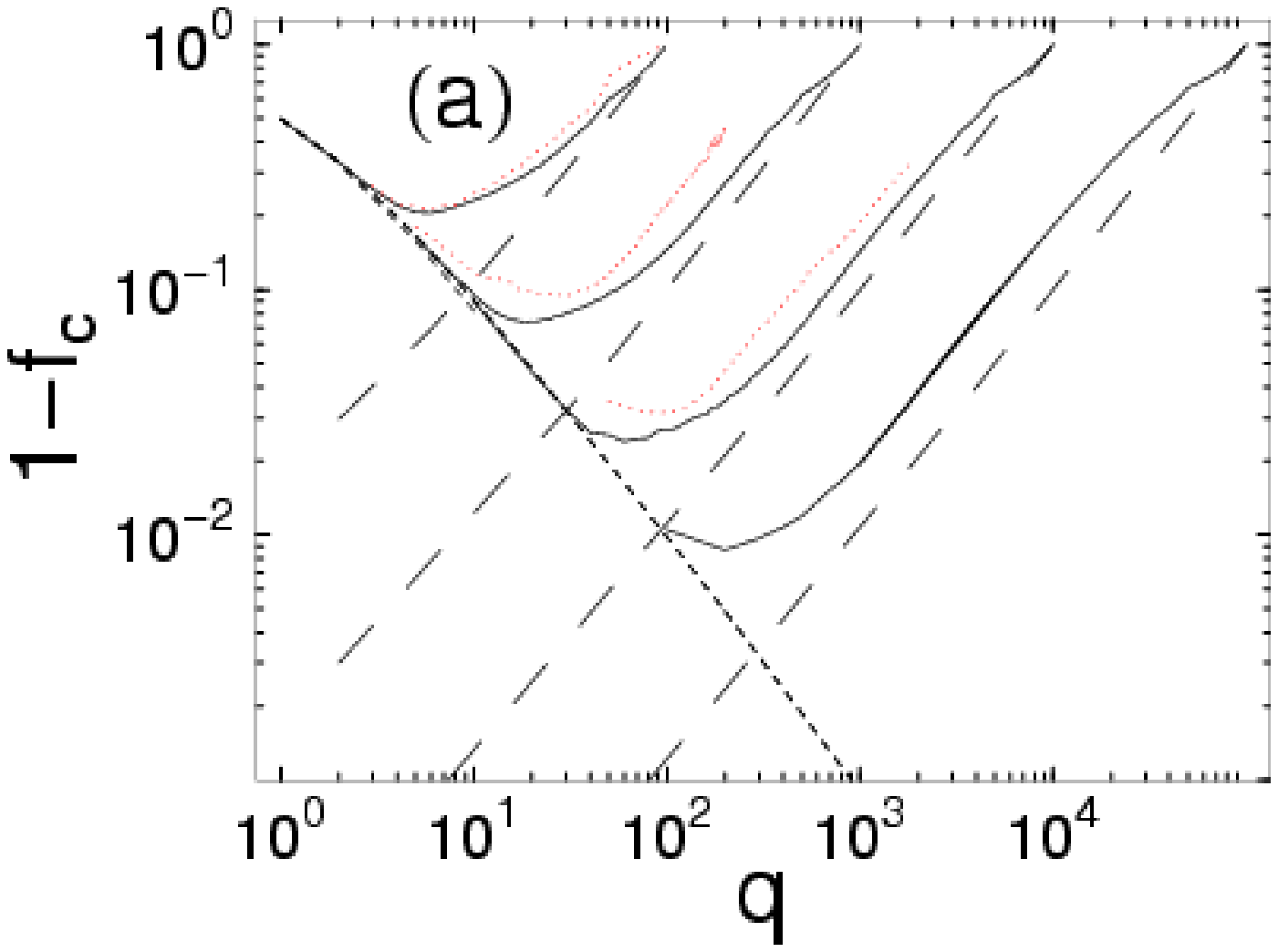}
}

\centerline{
\xsize
\epsfclipon
\epsfbox{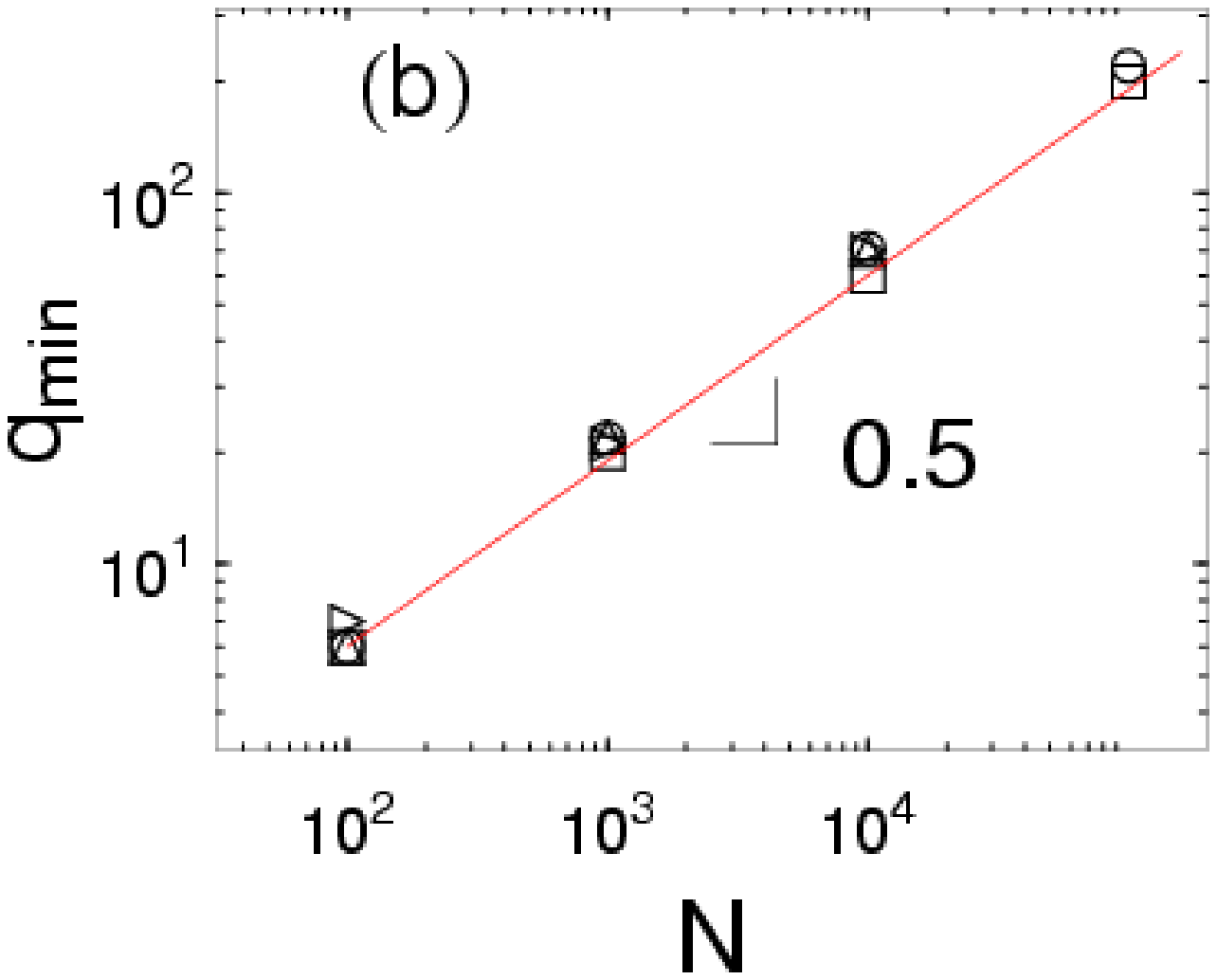}
}

\centerline{
\xsize
\epsfclipon
\epsfbox{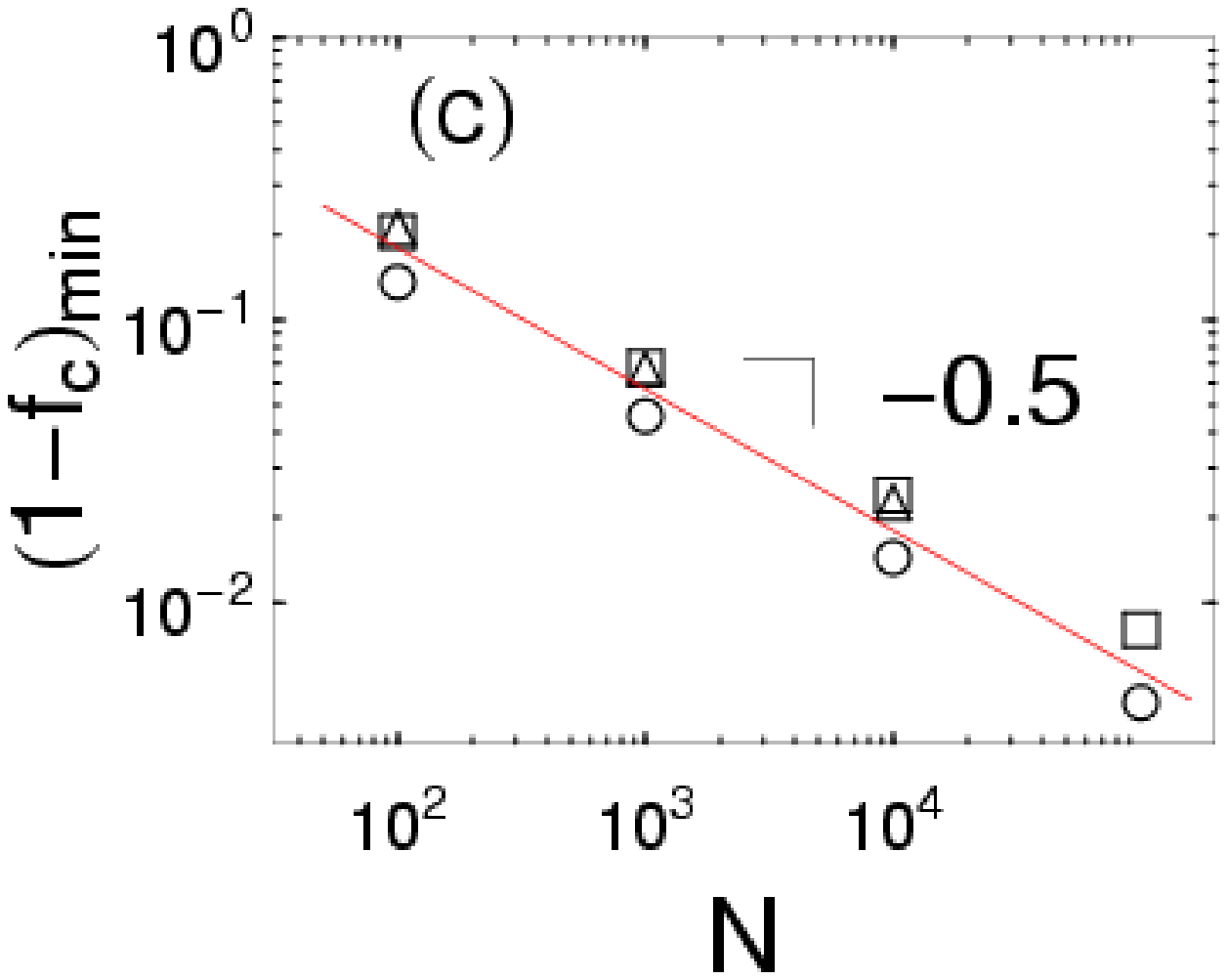}
}
\caption{For $\kav=2$ and for (from left to right) $N=10^2,10^3,10^4$
  and for random networks only $10^5$ (a) $1-f_c$ vs. number of hubs
  $q$.  The solid and dotted lines represent uncorrelated and correlated
  networks respectively.  Dashed lines(short) are approximation
  $f_c^{low}$; dashed lines(long) are approximation $f_c^{high}$.
  (b) Values of $q$, $q_{\rm min}$, at which $1-f_c$ is minimal vs. $N$.
  Squares and triangles represent uncorrelated and correlated networks
  respectively; circles represent $q^*$ the value of $q$ at
  which the approximations $f_c^{high}$ and $f_c^{low}$
  intersect. (c) Minimum values of $1-f_c$ versus $N$.  Squares and
  triangles represent uncorrelated and correlated networks respectively;
  circles represent the values of $(1-f_c)$ at $q=q^*$. }
\label{kav2}
\end{figure}

In Fig.~\ref{kav2}(a), we plot $1-f_c$ versus $q$ for $N=10^2,10^3,10^4$
and $10^5$ and $\kav=2$.  In Fig. 2(b) we plot the location of the
minima $q_{\rm min}$ versus $N$.  As expected $q_{\rm min}$ scales with $N$ as
$N^{0.5}$ and as shown in Fig. 2(c) the minimum values of $1-f_c$ scale
as $N^{-0.5}$.

Also shown in Fig.~\ref{kav2}(a) are plots for approximations to
$f_c$, $f_c^{\rm high}$ and $f_c^{\rm low}$, which we expect to be valid
respectively for high and low values of $q$.  We will use these
approximations as another way to show that $q_{\rm min}$ and $(1-f_c)_{\rm min}$
scale as found above.  The approximations are determined as follows:

\begin{itemize}

\item[{(i)}] When $q \sim N$ (i.e., the network is homogeneous), we
expect Eq.~(\ref{fc}) to hold, so $f_c^{\rm high}=1-1/(\kappa-1)$.  For
general $\kav$, using the distribution in Eqs.~(\ref{pkx}), we find for
$N \gg q \gg 1$
\begin{equation}
f_c^{\rm high}=1-{q \over{(\kav-1)N}}.
\label{fchigh}
\end{equation}
\item[{(ii)}] As found in \cite{Paul2005}, Eq.~(\ref{fc}) is not valid
for small $q$.  We must use an approximation based on the
fact that for small $q$ the network loses global connectivity when all
$q$ high degree nodes are removed. To first order in $1/q$
\cite{Paul2005}
\begin{equation}
\label{fclow3}
1-f_c^{\rm low}=1-{1 \over{q}}.
\end{equation}

\end{itemize}
Equating Eqs.~ (\ref{fchigh}) and  (\ref{fclow3}) we find the value of $q$
at which the approximations intersect
\begin{equation}
\label{qintersect}
q^*=\sqrt{\kav-1} \sqrt{N}.
\end{equation}
From the fact that $q*$ scales like $\sqrt{N}$, we conclude that all
characteristic values including the location of the minimum of (1-$f_c$)
scale like $\sqrt{N}$ with a prefactor dependent on $\kav$.

From Eqs.~(\ref{fclow3})and (\ref{qintersect}) we find for large $N$,
\begin{equation}
\label{vintersect}
1-f_c^{*}={1 \over{\sqrt{\kav-1}}} {1 \over \sqrt{N}  }.
\end{equation}
where $f_c^{*}$ is the value of value of $f_c$ where the approximations
intersect.  The scaling of $q^*$ and $1-f_c*$ are shown in
Figs.~\ref{kav2}(b) and (c).

\begin{figure}
\centerline{
\xsize
\epsfclipon
\epsfbox{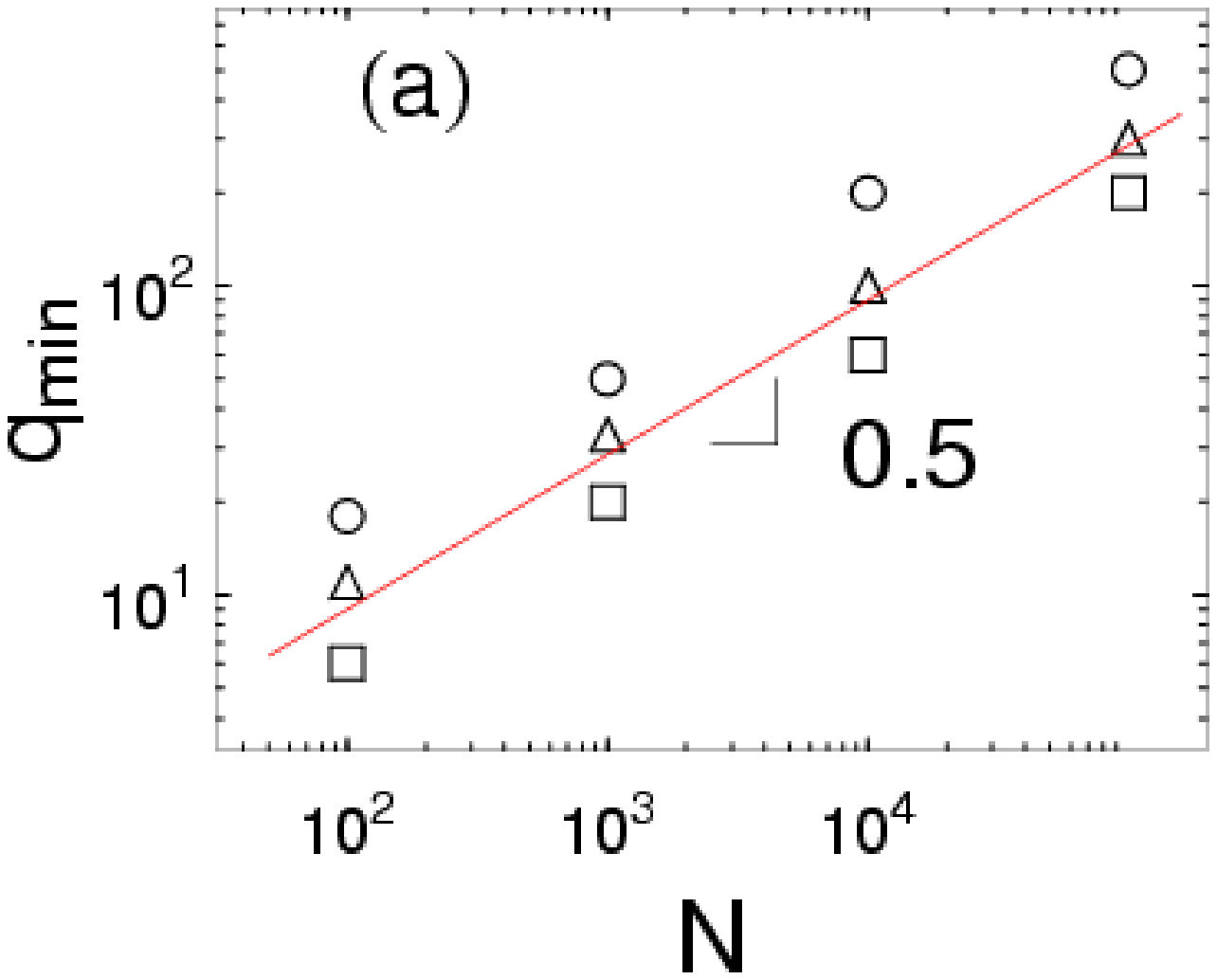}
}

\centerline{
\xsize
\epsfclipon
\epsfbox{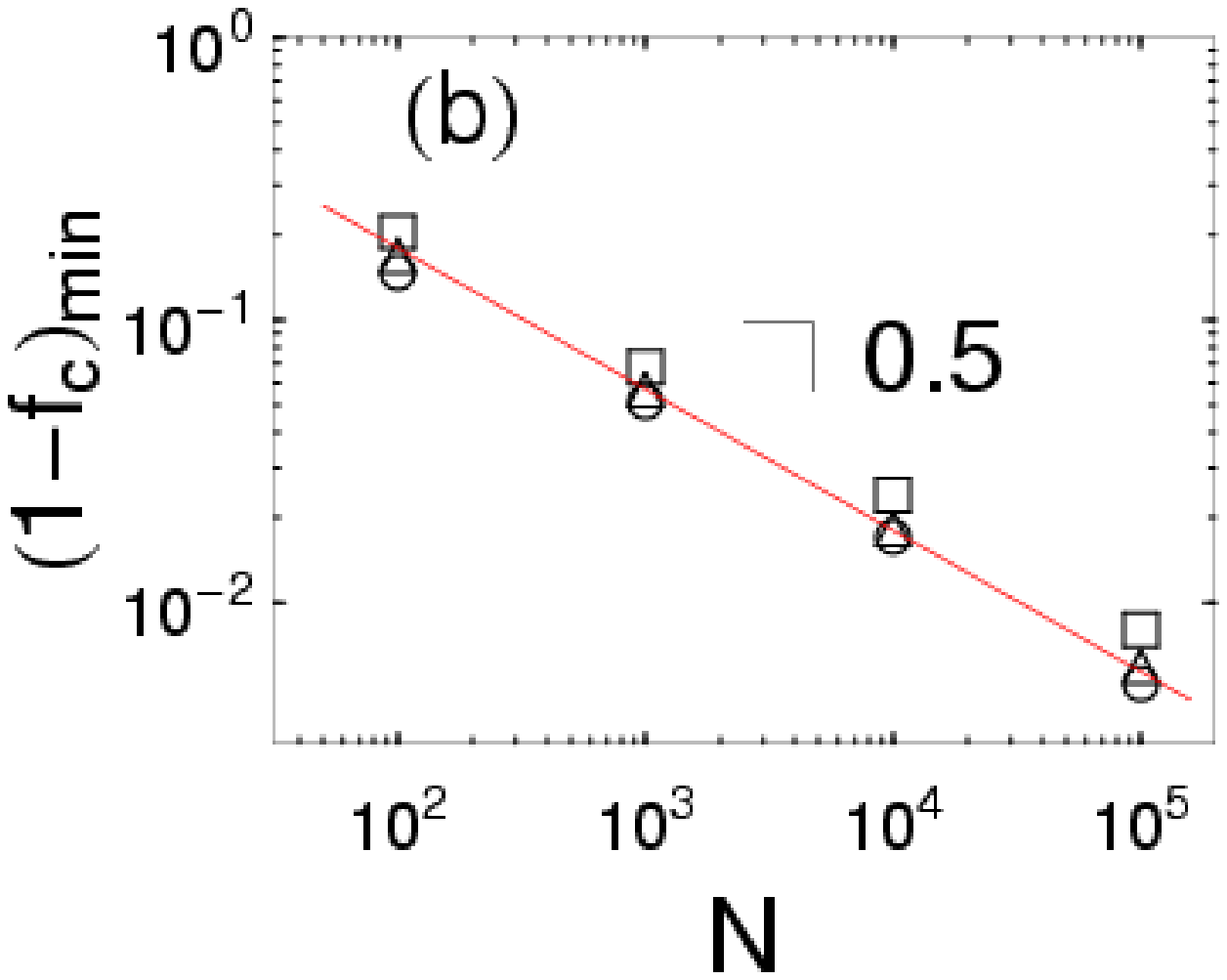}
}
 \caption{(a) Values of $q$, $q_{\rm min}$, at which $1-f_c$ is minimal vs. $N$.
  Squares, triangles and circles represent network with $\kav=2,3,$ and
  4 respectively (b) Minimum values of $1-f_c$ versus $N$. Squares,
  triangles and circles represent networks with $\kav=2,3,$ and 4
  respectively}
\label{comb234}
\end{figure}

We next study the effect of changing $\kav$.  Figs.~\ref{comb234}(a) and
(b) contain plots of $q_{\rm min}$ and $(1-f_c)_{\rm min}$ respectively for
$\kav=2,3,$ and 4.  We note that the scaling is independent of $\kav$
with only a change in the prefactor.

\section{Correlated Networks}

\begin{figure}
\centerline{
\xsize
\epsfclipon
\epsfbox{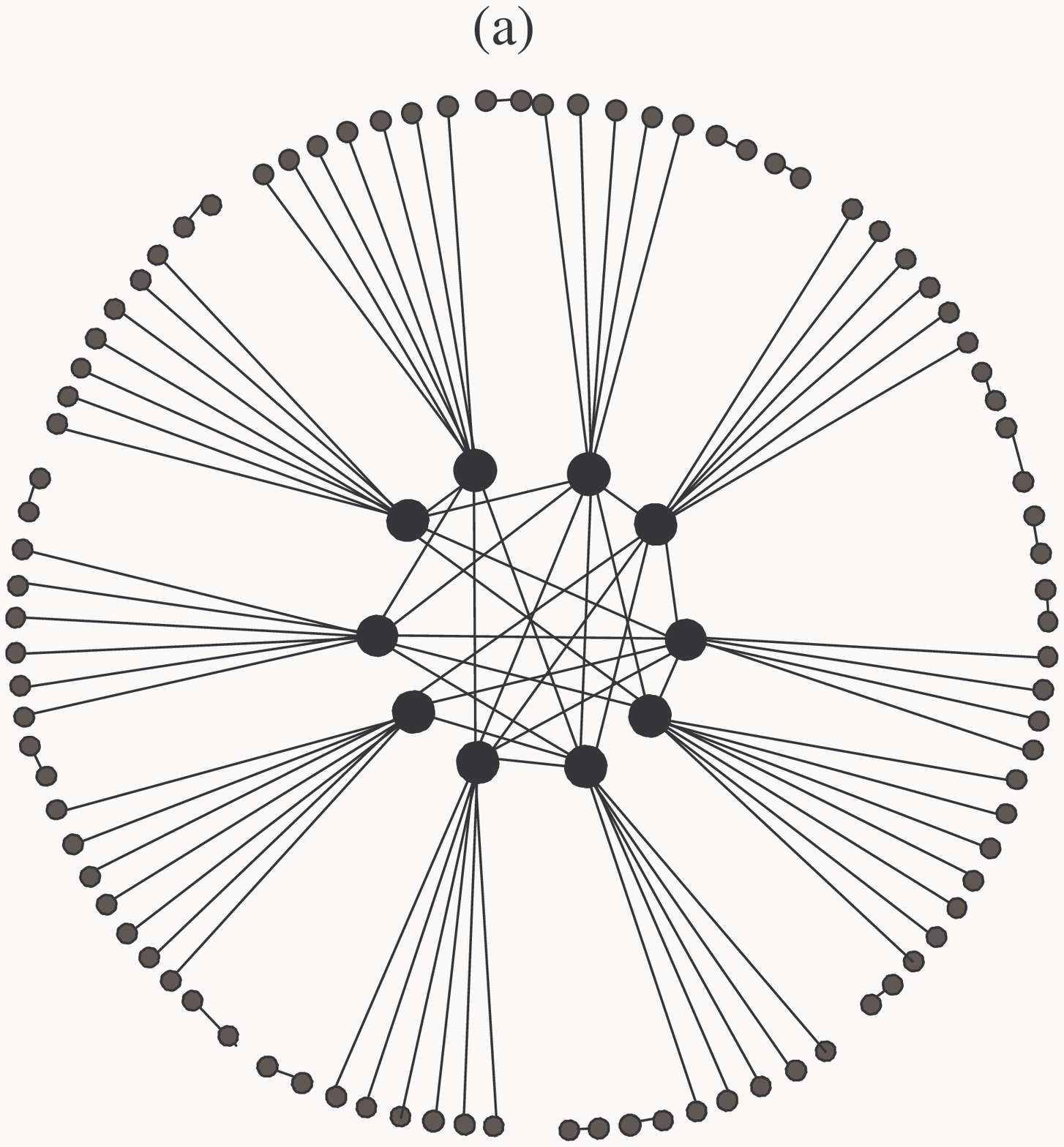}
}

\centerline{
\xsize
\epsfclipon
\epsfbox{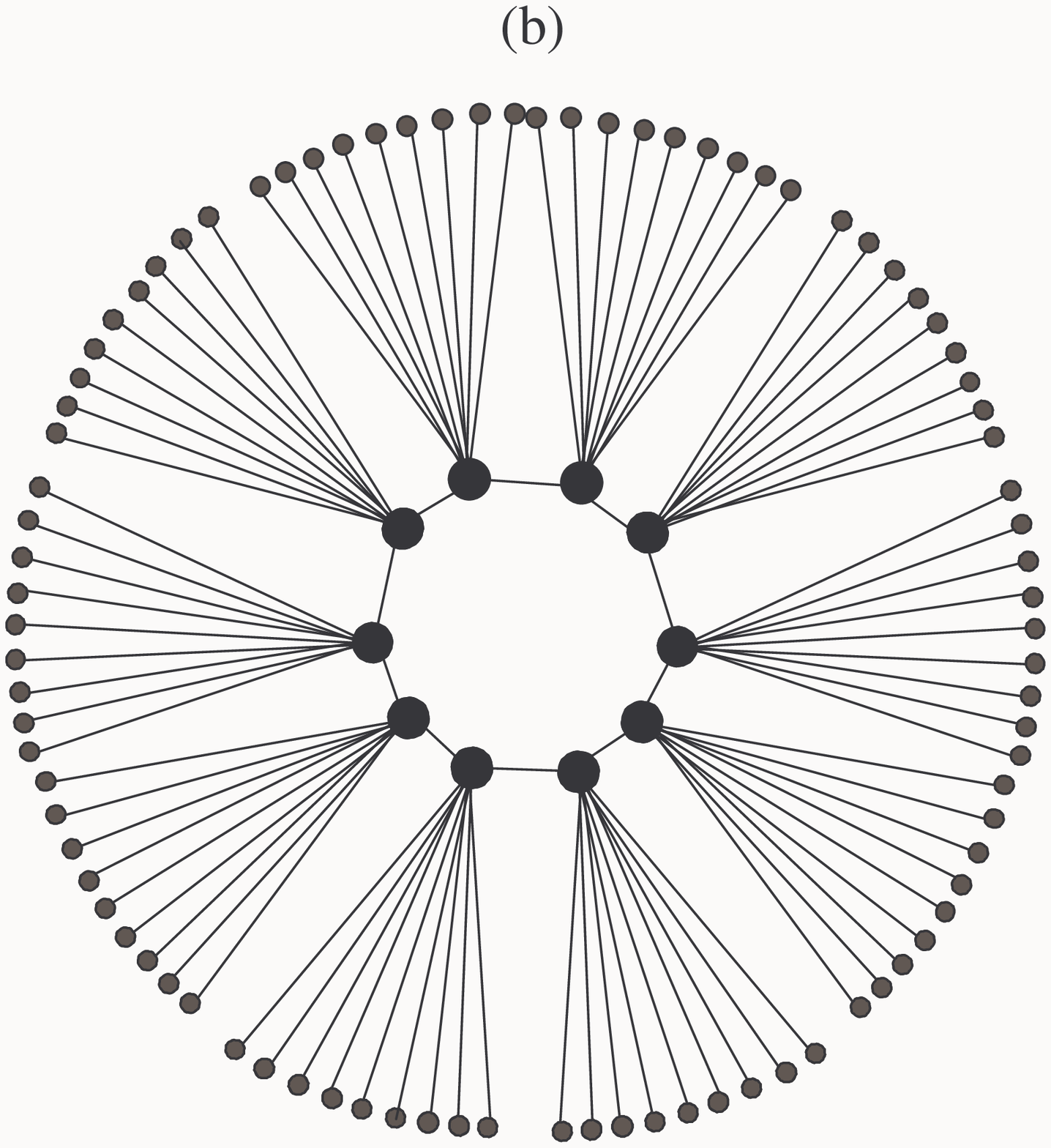}
}
\caption{Examples of 100 node networks with degree distribution given by
 Eqs.~(\ref{pkx}) with $\kav=2$.  (a) uncorrelated network.  Note that
 there are disconnected pairs of nodes of degree 1.  (b) correlated
 network in which each degree 1 node is connected to a high degree node.
 }
\label{example}
\end{figure}

In Fig.~\ref{example}(a) we show an example of a randomly created graph.
Note that, because the graph is created randomly, there are some
disjoint portions of the graph consisting of pairs of nodes connected to
each other.  Thus the network does not consist of a single connected
component.  We now study correlated networks which do not have this
shortcoming by disallowing connections between degree one nodes so that
the resulting network is a single cluster (see Fig.~\ref{example}(b)
which has the same degree distribution as Fig.~\ref{example}(a) ).

For correlated networks, the criteria for network collapse is
\cite{Newman2002}
\begin{equation}
\label{detA}
det({\bf A})=0
\end{equation}
where {\bf A} is a matrix containing elements $A_{j,k}=k e_{j,k} + q_j
\delta_{i,j}$ with $e_{j,k}$ the joint probability of the remaining
degrees \cite{remaining}  of the two vertices at either end of a randomly
chosen edges and with $q_k$ the probability of the remaining degree of a
single vertex at the end of a randomly chosen edge.

We create networks having the degree distribution of Eq.~(\ref{pkx})
with $\kav=2$ but with the constraint that leaf nodes cannot be
connected to each other.  We proceed as for uncorrelated networks except
that after removal of an edge instead of calculating $\kappa$ we
calculate $det({\bf A})$ and note the number of nodes removed before
$det({\bf A})=0$.

In Fig.~\ref{kav2}(a) we plot $1-f_c$ versus $q$ for $N=10^2,10^3$ and
$10^4$ \cite{notex}.  We note that the plots are similar to but slightly
higher than the corresponding plots for the random networks. In
Fig.~\ref{kav2}(b) we plot the values of $q$ at which $1-f_c$ is minimal
and see that they scale in a manner similar to the scaling of the
positions of the minima for the random networks.

\section{Comparison with Scale-Free Networks}

Scale-free networks with $\lambda < 3$ are known to be very robust
against random attack \cite{Albert,Cohen2000} with $1-f_c$ approaching zero as
$N\to\infty$.  Here, we determine the large $N$ behavior of
$1-f_c$ for scale-free networks for a given value of $\kav$ and compare
the behavior with that of the optimal bimodal network.

We consider a scale-free degree distribution $P(k) \sim k^{-\lambda}$
with $m \le k \le K$.  For large $K$ and $2< \lambda <3$
\cite{Cohen2000},
\begin{equation}
\label{kappa0sf}
\kappa={2-\lambda \over{3-\lambda}} m^{\lambda-2} K^{3-\lambda}.
\end{equation}
Substituting in Eq.~(\ref{fc}) and setting $K=K_s$ we find that for large $K$
\begin{equation}
\label{fcsf}
1-f_c \sim K^{3-\lambda} \sim N^{(\lambda-3) \over{2}}.
\end{equation}
Only in the limit of $\lambda$ approaching 2, does $1-f_c \sim N^{-0.5}$
similar to Eq.~(\ref{fcminN}).
For $\lambda<2$,
\begin{equation}
\label{kappa0sfx}
\kappa=\frac{2-\lambda}{3-\lambda} K \sim \sqrt{N}.
\end{equation}
and
\begin{equation}
\label{fcsfx}
1-f_c \sim \frac{1}{\sqrt{N} }.
\end{equation}
but for $\lambda \le 2$, $\kav$ diverges with increasing K.  Thus for a
given value of $\kav$, $1-f_c$ for the optimal bimodal network always
approaches 0 faster than the optimal scale-free network \cite{sf}.

\begin{figure}
\centerline{
\xsize
\epsfclipon
\epsfbox{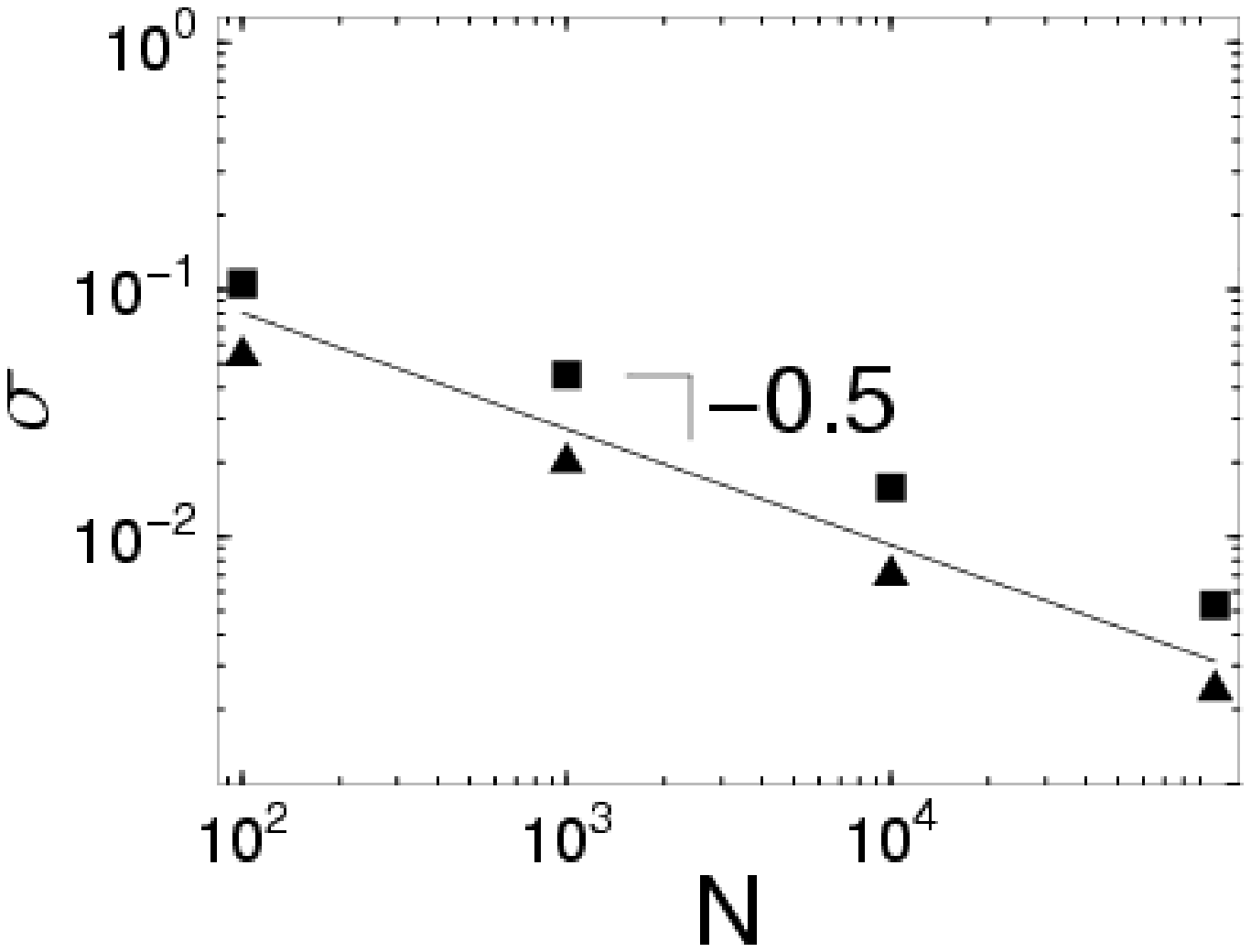}
}
\caption{Standard deviation in $f_c$ vs. $N$.  Squares represent optimal
bimodal configuration for $\kav=2$.  Triangles represent scale-free
configuration with $\lambda=2$}.
\label{pdevcomp}
\end{figure}

For completeness, to ensure that large variance is not a deficiency of
the optimal network, we now study how the variance in $f_c$ of the
optimal network compares with the variance of the scale-free networks .
Specifically in Fig.~\ref{pdevcomp} we plot the standard deviation
\begin{equation}
\sigma=\frac{\sqrt{ \langle (n_r -\langle n_r \rangle )^2 \rangle}}{N}
\label{sigma}
\end{equation}
vs $N$ for the optimal bimodal network with $\kav=2$ and for a
scale-free network with $\lambda=2$. For the scale-free network with
$\lambda=2$, $1-f_c \sim N^{-0.5}$ although it has a large value
of $\kav$.  We see that the standard deviation of $f_c$ of both networks
decreases as $N^{-0.5}$ with the scale-free network having a somewhat
smaller prefactor than the optimal network.  Thus the variance of $f_c$
is not a deficiency of the optimal network.

\section{Heuristic Argument for Optimal Configuration}

We now provide a heuristic argument for the optimal configuration which
applies to random or correlated networks. As shown above, the
configuration consists of $q \sim \sqrt{N}$ high degree nodes (hubs) of
degree $\sqrt{\kav N}$ and $N-q$ nodes of degree 1.  Intuitively, we
suspect that the optimal configuration is one in which there are many
leaf nodes with degree 1 connected to a network core composed of a much
smaller number of highly connected hubs node.  Removing a leaf node has
only a minimal impact on the connectivity of the network while removing
a hub has a much greater impact---but is much less probable.  It is not
obvious, however, how many hubs there should be. One might initially
suppose that the most robust network would be a single hub node
connected to all the remaining nodes (a {\it star} network).  It is easy
to show \cite{Paul2005}, however, that $f_c=1/2$ for this network which
is far from optimal.  To determine the number of hubs we proceed as
follows.  Consider first that there are $\kav N$ connections available
to construct the network.  Let $q$ denote the number of hubs. The number
of connections needed to connect the hubs to the leaf nodes is $2(N-q)$.
If we then make the argument that we want the hubs to form a complete
graph using the remaining connections we have
\begin{equation}
 q(q-1)=\kav N -2(N-q).
\label{complete}
\end{equation}
Solving for $q$ for large $N$ we have
\begin{equation}
q \sim \sqrt{\kav-2}\cdot\sqrt{N}
\label{completeN}
\end{equation}
and we again find that the number of hubs scales as $\sqrt{N}$ in a
manner similar to that implied by Eq.~(\ref{Pk2}), 
\begin{equation}               
q=\left({\kav-1\over\sqrt\kav}\right)\sqrt N,
\end{equation} 
for the optimal network with a different prefactor.

\section{Discussion and Summary}

We have shown analytically and confirmed numerically using Monte Carlo
simulations that networks with bimodal degree distributions, with
$q \sim \sqrt{N}$ high degree nodes (hubs) and $N-q$ nodes of degree 1, are
most robust to random breakdown.  Also we have shown that $1-f_c$
approaches $0$ as $1/\sqrt N$, faster than any other network
configuration including scale free networks. Finally, we have offered a
simple heuristic argument which explains these results.

\section{Acknowledgment}
We thank ONR for support.

\begin{thebibliography}{99}
\bibitem{notex}Because of the computation complexity of finding the
determinant of {\bf A} we do not treat $N=10^5$ as we did for random networks.
\bibitem{sf} We have also performed simulations that show that even when
$K > K_s$, $1-f_c$ does not approach $0$ faster than $1/ \sqrt{N}$ for any
value of $\lambda$.
\end{thebibliography}
\end{document}